\documentclass[%
 reprint,
superscriptaddress,
 amsmath,amssymb,
 aps,
 prl,
]{revtex4-1}

\usepackage{graphicx}
\usepackage{epstopdf}
\usepackage{xcolor}
\usepackage{dcolumn}
\usepackage{bm}
\usepackage{hyperref}

\begin{document}

\preprint{APS/123-QED}

\title{Additivity of R\"{o}ntgen term and recoil-induced correction to spontaneous emission rate}

\author{Anwei Zhang}
\email{hunnuzaw@163.com}
\affiliation{School of Physics and Astronomy, Shanghai Jiao Tong University, Shanghai 200240, China}

\author{Danying Yu}
\affiliation{State Key Laboratory of Advanced Optical Communication Systems and Networks, School of Physics and Astronomy, Shanghai Jiao Tong University, Shanghai 200240, China}

\begin{abstract}
For a moving atom, the spontaneous emission rate is modified due to the contributions from two factors, the R\"{o}ntgen interaction term and the recoil effect induced by the emitted photon. Here we investigate the emission rate of a uniformly moving atom near a perfectly conducting plate and obtain the corrections induced by these two factors.  We
 find that the corrections individually induced by the R\"{o}ntgen term and the recoil effect
 can be simply added and result in the total correction to the decay rate. Moreover, it is shown that the R\"{o}ntgen term gives positive correction, while the recoil effect induces negative correction. Our work paves the way towards the future studies of the light-matter interaction for the moving particle in quantum optics.
\end{abstract}

\maketitle


The spontaneous emission of an atom is of the
 fundamental importance in atomic physics and quantum optics \cite{book}.
Such a quantum process has been studied extensively  with atoms in various physical systems, including
cavity \cite{2,3}, photonic crystal \cite{4,5}, and metamaterial \cite{6}.
However, the atom in most of previous investigations is usually treated to be stationary or fixed.
Under this treatment, the conservation of momentum in the process of emitting or absorbing photon by the atom is neglected.
In order to describe the spontaneous emission accurately, one should take the motion of atom into account.

For the spontaneous decay from an atom involving nonrelativistic motion, one needs to consider two additional factors.
One is the R\"{o}ntgen interaction term \cite{c1, c2, c30, c3}, which describes the coupling between the motion of
the atom, the electric dipole moment, and the magnetic component of the radiation field.
The inclusion of the R\"{o}ntgen interaction term can successfully overcome the violation of the gauge invariance of radiation-reduced mechanical forces and the energy-momentum conservation from the atom-field Hamiltonian under the dipole approximation \cite{c1}.  The other one is the recoil shift \cite{15}, induced by the radiation of the photon which modifies the momentum and energy of the atom. It makes the atom  no longer resonant with the emitted photon, which, on the other hand, is the case for the stationary atom. Actually, the recoil shift is an addition to the Doppler shift and  has been measured at the order of kHz with the saturated absorption spectroscopy \cite{hall}.

 In this paper, we study the spontaneous decay of a uniformly moving atom near a perfectly conducting plate as a boundary, where corrections from both the R\"{o}ntgen term and the recoil shift are introduced. We find that the total correction of the spontaneous emission rate in this system is equivalent to the summation of corrections induced by involving the R\"{o}ntgen term and the recoil shift separately. Moreover,
  the R\"{o}ntgen term brings positive correction, while the recoil gives the negative correction. We discuss decay rates for such moving atom with dipoles at different directions. Our work reveals spontaneous emission details of the moving atom near the boundary, and therefore provides important insights into understanding this piece of the physical nature in quantum optics.

\begin{figure}[htbp]
\centering
\includegraphics[width=0.367\textwidth ]{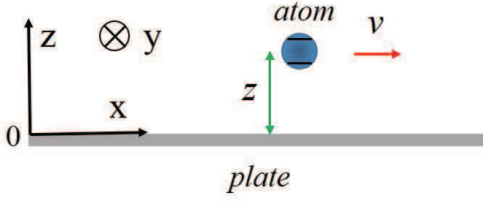}
\caption{Schematic illustration for a two-level atom moving at a constant velocity $v$ parallel to an infinitely perfectly conducting plate.}\label{figure.0}
\end{figure}

We start with considering a two-level atom with transition frequency $\omega_{0}$ and mass $m$ moving near an infinitely perfectly conducting plate located at $z=0$ (see Fig.~\ref{figure.0}). The total Hamiltonian of our system under nonrelativistic condition
can be described by $H=H_0+H_I$, where  $H_0$ gives the atomic centre-of-mass kinetic energy:
\begin{equation}\label{1}
 H_0=\frac{\hat{\textbf{P}}^{2}}{2m}+\omega_0\hat{\sigma}^{+}\hat{\sigma}^{-}+\sum_{\textbf{k}\lambda}\omega_{\textbf{k}} \hat{a}^{\dag}_{\textbf{k}\lambda}\hat{a}_{\textbf{k}\lambda}.
\end{equation}
Here $\hat{\textbf{P}}$ denotes the canonical momentum and $\hat{\sigma}^{+}=|e\rangle\langle g|$ ($\hat{\sigma}^{-}=|g\rangle\langle e|$) is the raising (lowering) operator for the atomic transition. The last term in Eq.~(\ref{1}) describes the
  quantized electromagnetic field with the creation operators $\hat{a}^{\dag}_{\textbf{k}\lambda}$ and annihilation operators $\hat{a}_{\textbf{k}\lambda}$ for a photon with the wave vector $\textbf{k}$,
the frequency $\omega_{\textbf{k}}$, and the polarization $\lambda=1,2$. We use the natural units $\hbar=c=1$ throughout the text for the simplicity.
The atom-field interaction Hamiltonian $H_I$ is given by \cite{c2,c3}
\begin{equation}\label{2}
H_I=-\hat{\textbf{d}}\cdot \hat{\textbf{E}}(\textbf{r})-\frac{1}{2m}\{\hat{\textbf{P}}\cdot [\hat{\textbf{B}}(\textbf{r})\times \hat{\textbf{d}}]+[\hat{\textbf{B}}(\textbf{r})\times \hat{\textbf{d}}]\cdot \hat{\textbf{P}}\},
\end{equation}
 where $\hat{\textbf{d}}=\textbf{d}(\hat{\sigma}^{+}+\hat{\sigma}^{-})$ denotes the atomic dipole operator with dipole transition moment $\textbf{d}$, and $\hat{\textbf{E}}(\textbf{r})$, $\hat{\textbf{B}}(\textbf{r})$ are the electric field and magnetic field operators at the position $\textbf{r}$ of the atom, respectively. The interaction described by Eq.~(\ref{2}) includes the R\"{o}ntgen term which results from the motion of the atom, i.e., the second term.

 In the interaction picture, the effective interaction Hamiltonian can be written as
\begin{eqnarray}\label{3}
H_I(t)=&-&(\hat{\sigma}^{+}e^{i\omega_{0}t}+\hat{\sigma}^{-}e^{-i\omega_{0}t})\bigg\{\textbf{d}\cdot \hat{\textbf{E}}(\textbf{r},t)\nonumber \\
&+&\frac{1}{2m}\{\hat{\textbf{P}}\cdot [\hat{\textbf{B}}(\textbf{r},t)\times \textbf{d}]
+[\hat{\textbf{B}}(\textbf{r},t)\times \textbf{d}]\cdot \hat{\textbf{P}}\}\bigg\}.\nonumber \\
\end{eqnarray}

In the presence of the infinite perfectly conducting plate placed at $z=0$, the mode functions of vector potential $\textbf{A}$ can be written as \cite{book}
\begin{eqnarray}\label{4}
  (A_{\textbf{k}\lambda})_{x} &=& i\frac{\varepsilon_{k}}{\omega_{\textbf{k}}} (e_{\textbf{k}\lambda})_{x}e^{i\textbf{k}_{\parallel}\textbf{r}-i\omega_{\textbf{k}}t}\sin(k_{z}z), \nonumber\\
  (A_{\textbf{k}\lambda})_{y} &=& i\frac{\varepsilon_{k}}{\omega_{\textbf{k}}} (e_{\textbf{k}\lambda})_{y}e^{i\textbf{k}_{\parallel}\textbf{r}-i\omega_{\textbf{k}}t}\sin(k_{z}z), \nonumber\\
  (A_{\textbf{k}\lambda})_{z} &=&\frac{\varepsilon_{k}}{\omega_{\textbf{k}}} (e_{\textbf{k}\lambda})_{z}e^{i\textbf{k}_{\parallel}\textbf{r}-i\omega_{\textbf{k}}t}
  \cos(k_{z}z),
\end{eqnarray}
where $\varepsilon_{k}=\sqrt{\omega_{\textbf{k}}/(\varepsilon_{0}V})$, $\varepsilon_{0}$ is the permittivity of vacuum, $V$ is the quantization volume, $\textbf{e}_{\textbf{k}1}=i(-k_{y}/k_{\parallel},k_{x}/k_{\parallel},0)$, $\textbf{e}_{\textbf{k}2}=[-k_{x}k_{z}/(k_{\parallel}\omega_{\textbf{k}}),
-k_{y}k_{z}/(k_{\parallel}\omega_{\textbf{k}}),k_{\parallel}/\omega_{\textbf{k}}]$, and $\textbf{k}_{\parallel}=(k_{x},k_{y},0)$ with $k_{\parallel}=\sqrt{k^{2}_{x}+k^{2}_{y}}$. It can be verified that the mode functions in Eq.~(\ref{4}) satisfy the boundary condition, the orthogonality condition, and the Coulomb gauge condition. Thus we obtain the negative-frequency components of the electric field and magnetic field operators
\begin{eqnarray}\label{5}
  \hat{E}^{-}_{i}(\textbf{r},t) &=&-i \sum_{\textbf{k}\lambda} \omega_{\textbf{k}}(A_{\textbf{k}\lambda})^{\ast}_{i}\hat{a}^{\dag}_{\textbf{k}\lambda}, \nonumber\\
  \hat{B}^{-}_{x}(\textbf{r},t) &=& -i \sum_{\textbf{k}\lambda}\varepsilon_{k}(\textbf{h}\times \textbf{e}_{\textbf{k}\lambda})^{\ast}_{x}e^{-i\textbf{k}_{\parallel}\textbf{r}+i\omega_{\textbf{k}}t}
  \cos(k_{z}z)\hat{a}^{\dag}_{\textbf{k}\lambda},\nonumber\\
 \hat{B}^{-}_{y}(\textbf{r},t) &=& -i \sum_{\textbf{k}\lambda}\varepsilon_{k}(\textbf{h}\times \textbf{e}_{\textbf{k}\lambda})^{\ast}_{y}e^{-i\textbf{k}_{\parallel}\textbf{r}+i\omega_{\textbf{k}}t}
  \cos(k_{z}z)\hat{a}^{\dag}_{\textbf{k}\lambda},\nonumber\\
  \hat{B}^{-}_{z}(\textbf{r},t) &=&- \sum_{\textbf{k}\lambda}\varepsilon_{k}(\textbf{h}\times \textbf{e}_{\textbf{k}\lambda})^{\ast}_{z}e^{-i\textbf{k}_{\parallel}\textbf{r}+i\omega_{\textbf{k}}t}
  \sin(k_{z}z)\hat{a}^{\dag}_{\textbf{k}\lambda}.\nonumber\\
\end{eqnarray}
Here $\textbf{h}=\textbf{k}/\omega_{\textbf{k}}$ is the unit vector of wave vector. The positive-frequency components are the Hermitian conjugate of the corresponding negative-frequency components.

We consider initially the atom prepared at the excited state, and moving with the canonical momentum state $|\textbf{p}\rangle$, while the field is in vacuum state. Therefore, we can write the initial state of the system as $|i\rangle=|e,\textbf{p},0\rangle$.
Due to the term $\hat{\sigma}^{-}\hat{a}^{\dag}_{\textbf{k}\lambda}$ in the interaction Hamiltonian, the state of the moving atom evolves into its ground state by emitting a photon, which gives the final state of the system $|f\rangle=|g,\textbf{p}-\textbf{k},1_{\textbf{k}}\rangle$. The derivative of the probability of the transition under the first-order approximation gives the spontaneous decay rate of the atom:
\begin{eqnarray}\label{6}
\Gamma&=&\partial_{t}\sum_{\textbf{k}\lambda}\mid\int^{t}_{0}d t^{\prime}\langle f|H_I(t^{\prime})|i\rangle\mid^{2}\nonumber\\
&=&2\mathrm{Re}\sum_{\textbf{k}\lambda}\int^{t}_{0}d t^{\prime}\langle i|H_I(t)|f\rangle\langle f|H_I(t^{\prime})|i\rangle.
\end{eqnarray}

In order to calculate the decay rate, we consider the case that the atom moves parallel to the plate, i.e., $\textbf{p}=(p,0,0)$. If we choose the dipole transition moment $\textbf{d}=(0,0,d)$, which exhibits the direction normal to the boundary, there is the equality: $\langle f|\hat{\textbf{P}}\cdot [\hat{\textbf{B}}(\textbf{r},t)\times \textbf{d}]
+[\hat{\textbf{B}}(\textbf{r},t)\times \textbf{d}]\cdot \hat{\textbf{P}}|i\rangle=d\langle f|2\hat{B}^{-}_{y}p-i\partial_{x}\hat{B}^{-}_{y}+i\partial_{y}\hat{B}^{-}_{x}|i\rangle=
d\langle f|(2p-k_{x})\hat{B}^{-}_{y}+k_{y}\hat{B}^{-}_{x}|i\rangle$. Here the formula $[\hat{P}_{i},\hat{B}_{j}]=-i\partial_{i}\hat{B}_{j}$ is used.
We then rewrite Eq.~(\ref{6}) to be
\begin{equation}\label{7}
  \Gamma  = 2\pi d^{2}\sum_{\textbf{k}\lambda}\varepsilon^{2}_{k}F^{(1)}_{\lambda}(\omega_{\textbf{k}})
  \delta(\omega_0-\omega_{\textbf{k}}+k_{x}v)\cos^{2}(k_{z}z),
\end{equation}
where $F^{(1)}_{\lambda}(\omega_{\textbf{k}})=|(e_{\textbf{k}\lambda})_{z}+(p-h_{x}\omega_{\textbf{k}}/2)(\textbf{h}\times \textbf{e}_{\textbf{k}\lambda})_{y}/m+h_{y}\omega_{\textbf{k}}(\textbf{h}\times \textbf{e}_{\textbf{k}\lambda})_{x}/(2m)|^{2}$ and $v(\ll 1)$ is the velocity of the atom. In the derivation, we have used the relation $x=v t$ and the Markov approximation.

For our atom with the nonrelativistic motion velocity $\textbf{v}$,
conservations of momentum and energy give  \cite{aa}
\begin{eqnarray}
  \textbf{k}+\Delta \textbf{p} &=& 0, \nonumber\\
  \Delta E_{a}+\Delta \textbf{p}\cdot\textbf{v}+\omega_{ \textbf{k}} &=& 0,
\end{eqnarray}
where $\Delta E_{a}=-\omega_{0}$ is the change of the atom's s internal energy and $\Delta \textbf{p}\cdot\textbf{v}=-\textbf{k}\cdot\textbf{v}=-k_{x}v$
describes the change of the atom's kinetic energy $(\textbf{p}-\textbf{k})^{2}/2m-\textbf{p}^{2}/2m=-(ph_{x}\omega_{\textbf{k}}-\omega^{2}_{\textbf{k}}/2)/m$.
Here $ph_{x}\omega_{\textbf{k}}/m$ and $\omega^{2}_{\textbf{k}}/(2m)$ denotes the the Doppler shift and the recoil shift \cite{c3,15}, respectively.
The $\delta$-function in Eq.~$({\ref{7}})$ reflects the conservation of energy, which is expected from Fermi's golden rule. Here the quantity $\textbf{v}$ is not equal to $\textbf{p}/m$ as a consequence of the difference between the kinetic momentum $m d\hat{\textbf{r}}/dt$ and the canonical momentum $\hat{\textbf{P}}$ \cite{c2}.

We now perform the continuum limit, i.e., $V\rightarrow\infty$, and obtain
\begin{equation}\label{8}
 \Gamma=\frac{d^{2}}{(2\pi)^{2}\varepsilon_{0}}\sum_{\lambda}\int d\Omega\frac{\omega^{3}_{+}F^{(1)}_{\lambda}(\omega_{+})}{G(\omega_{+})}
  \cos^{2}(h_{z}z\omega_{+}),
\end{equation}
where $d\Omega$ is the solid angle, $\omega_{+}=h_{x} p-m+\sqrt{\left(m-h_{x} p\right)^{2}+2 \omega_{0}m }$ is the positive root of the equality $\omega_0-\omega_{\textbf{k}}+k_{x}v=0$, and $G(\omega_{+})=\partial_{\omega_{\textbf{k}}}|\omega_0-
 \omega_{\textbf{k}}+k_{x}v|_{\omega_{+}}=1+(\omega_{+}-p h_{x})/m$.
We assume the atom is heavy enough, i.e., $m\gg p, \omega_{\textbf{k}}, \omega_{0}$. Then we can expand $F^{(1)}_{\lambda}(\omega_{+})$ to the first order of $1/m$ and sum the expanded expressions over the polarisations $\lambda$ to obtain,
\begin{equation}\label{9}
 \sum_{\lambda}F^{(1)}_{\lambda}(\omega_{+})=1-h^{2}_{z}+\left[\left(h^{2}_{x}+h^{2}_{y}\right)\omega_{+}
 -2ph_{x}\right]/m.
\end{equation}
Here the relation $\sum_{\lambda}(e_{\textbf{k}\lambda})_{i}(e_{\textbf{k}\lambda})^{\ast}_{j}=\delta_{ij}-h_{i}h_{j}$ is used.

 \begin{figure*}[htbp]
\centering
\includegraphics[width=0.317\textwidth ]{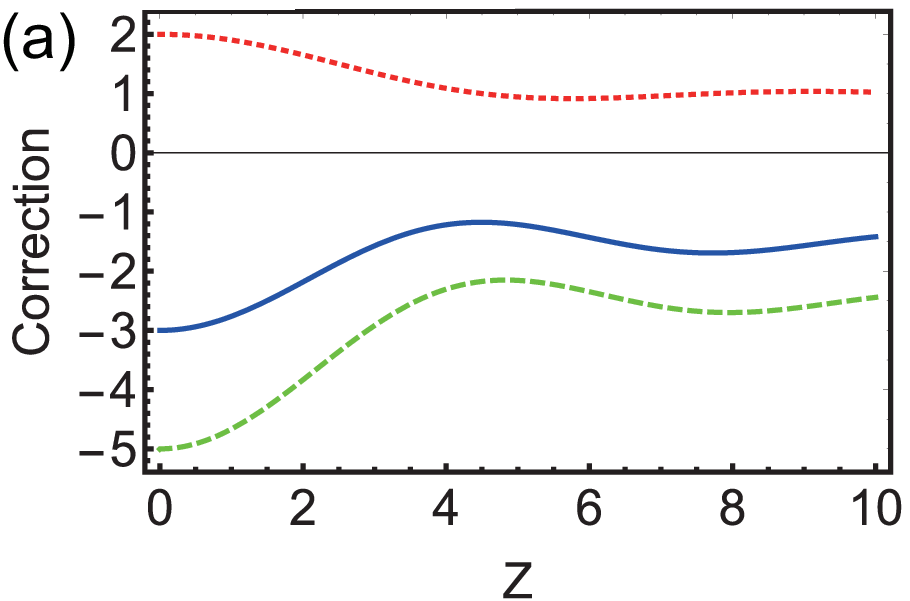}
\includegraphics[width=0.317\textwidth ]{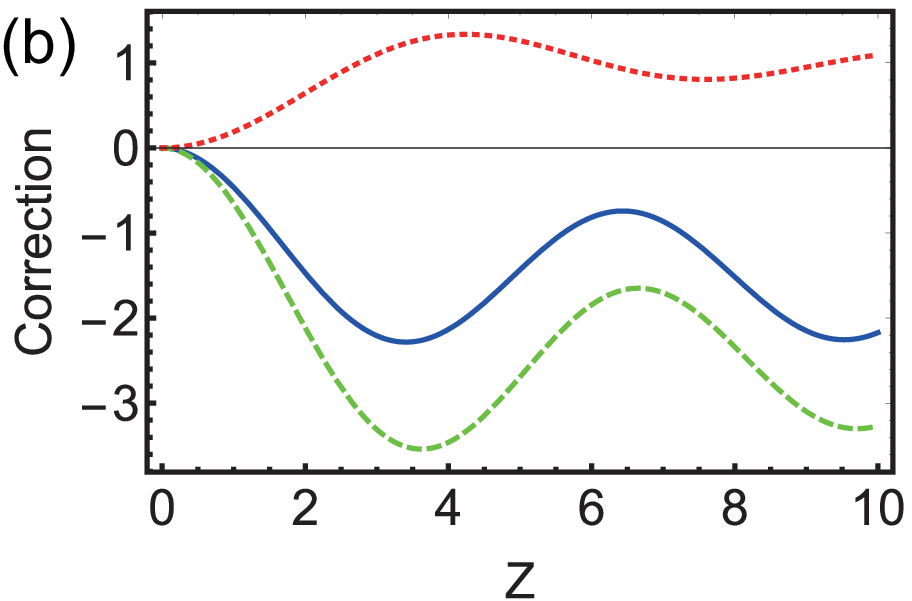}
\includegraphics[width=0.317\textwidth ]{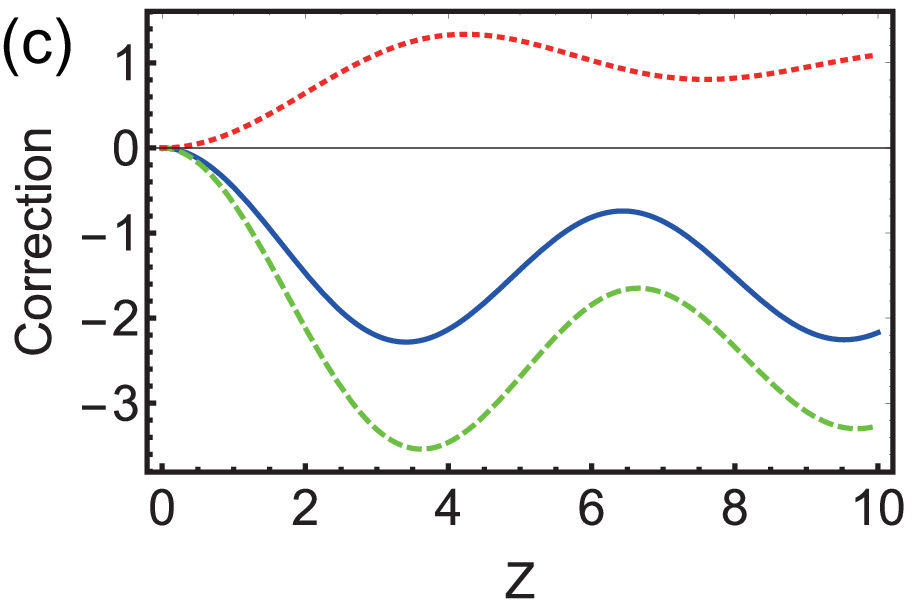}
\includegraphics[width=0.357\textwidth ]{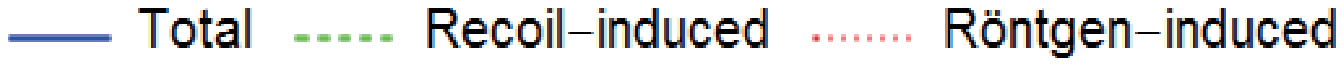}
\caption{The correction to the spontaneous emission rate in unit of $\Gamma_{0}\omega_{0}/m$ as a function of $Z$ for the dipole transition
moment in $z$, $y$, $x$-direction, respectively.}\label{figure.1}
\end{figure*}

By substituting Eq.~(\ref{9}) into Eq.~(\ref{8}) and then taking integration over the solid angle in the first order approximation of $1/m$, we derive the decay rate
\begin{equation}\label{10}
  \Gamma=\Gamma_{bz}-\frac{3\Gamma_{0}\omega_{0}}{2m}\bigg(1+\frac{\sin Z}{Z}\bigg),
\end{equation}
where
  $\Gamma_{bz}=\Gamma_{0}\left(1-3\frac{Z\cos Z-\sin Z}{Z^{3}}\right)$
gives the spontaneous decay rate of a fixed atom with $z$-direction dipole transition
moment in the presence of the boundary \cite{book}. Here $\Gamma_{0}=\omega^{3}_{0}d^{2}/(3\pi\varepsilon_{0})$ is the decay rate of an atom fixed in free space. $Z$ is a dimensionless parameter, which reads
 \begin{equation}\label{n2}
  Z=2 z \omega_{0}.
 \end{equation}
  The second term, i.e., $\Gamma_{ct} =-\frac{3\Gamma_{0}\omega_{0}}{2m}\left(1+\frac{\sin Z}{Z}\right)$, in Eq.~(\ref{10}) is the correction to the decay rate, which includes the contributions from both the R\"{o}ntgen term and the recoil of the emitted photon. It is interesting to notice that such contributions from the two terms hold the additivity. If there is only the R\"{o}ntgen term that is involved in the calculation, and with the same procedure, we find that the correction to the decay rate is
  \begin{equation}
   \Gamma_{cR}=\Gamma_{bz}\omega_{0}/m,
  \end{equation}
     which gives the positive contribution to the decay rate [see Fig.~\ref{figure.1}(a)]. Nevertheless,  if we only consider the recoil effect, the correction to the  decay rate is
      \begin{equation}
  \Gamma_{cr}=\frac{\Gamma_{0}\omega_{0}}{m}\left[-\frac{5}{2}+3\frac{2Z\cos Z-(Z^{2}+2)\sin Z}{2Z^{3}}\right],
  \end{equation}
        which gives the negative contribution to the decay rate. It can be found that $\Gamma_{ct} = \Gamma_{cR} +\Gamma_{cr}$, showing that the summation between the R\"{o}ntgen term induced correction and the recoil  effect induced correction leads to the total correction to the spontaneous decay rate.

Note that since the Eq.~(\ref{10}) is obtained by expanding the integrand in Eq.~(\ref{8}) to first order in $1/m$, there is the constraint $o(m^{2})Z^{2}/m^{2}\ll1$, i.e., $Z$ can not be extended to infinity and our approximate results are only valid when the atom is near the plate.

We have discussed the case of the moving atom near the boundary with the dipole transition moment along the $z$-direction. Next, we consider cases with the dipole along the $y$-direction and $x$-direction, separately.  By setting the dipole along the $y$-direction, i.e., $\textbf{d}=(0,d,0)$, we have the equality: $\langle f|\hat{\textbf{P}}\cdot [\hat{\textbf{B}}(\textbf{r},t)\times \textbf{d}]
+[\hat{\textbf{B}}(\textbf{r},t)\times \textbf{d}]\cdot \hat{\textbf{P}}|i\rangle=-d\langle f|2\hat{B}^{-}_{z}p-i\partial_{x}\hat{B}^{-}_{z}+i\partial_{z}\hat{B}^{-}_{x}|i\rangle$.
We therefore can derive the decay rate as
\begin{eqnarray}\label{11}
  \Gamma&=& 2\pi d^{2}\sum_{\textbf{k}\lambda}\varepsilon^{2}_{k}F^{(2)}_{\lambda}(\omega_{\textbf{k}})
  \delta(\omega_0-\omega_{\textbf{k}}+k_{x}v)\sin^{2}(k_{z}z)\nonumber\\
  &=&\Gamma_{by}-\frac{3\Gamma_{0}\omega_{0}}{2m}\bigg(1-\frac{\cos Z}{2}-\frac{\sin Z}{2Z}\bigg),
\end{eqnarray}
where $F^{(2)}_{\lambda}(\omega_{\textbf{k}})=|(e_{\textbf{k}\lambda})_{y}-(p-h_{x}
\omega_{\textbf{k}}/2)(\textbf{h}\times \textbf{e}_{\textbf{k}\lambda})_{z}/m-h_{z}
\omega_{\textbf{k}}(\textbf{h}\times \textbf{e}_{\textbf{k}\lambda})_{x}/(2m)|^{2}$ and
  $\Gamma_{by}=\Gamma_{0}\left[1-3\frac{Z\cos Z+(Z^{2}-1)\sin Z}{2Z^{3}}\right]$
gives the spontaneous decay rate of a fixed atom with the  dipole transition
moment along the $y$-direction in the presence of the plate \cite{book}. Here $\sum_{\lambda}F^{(2)}_{\lambda}(\omega_{\textbf{k}})=1-h^{2}_{y}+\left[\left(h^{2}_{x}+h^{2}_{z}\right)
\omega_{\textbf{k}}-2ph_{x}\right]/m$ is used.
The second term, i.e., $\Gamma_{ct}=-\frac{3\Gamma_{0}\omega_{0}}{2m}\left(1-\frac{\cos Z}{2}-\frac{\sin Z}{2Z}\right)$, in Eq.~(\ref{11}) is the correction to the spontaneous decay rate. If we only consider the R\"{o}ntgen term, it will be reduced to
\begin{equation}
 \Gamma_{cR}=\Gamma_{by}\omega_{0}/m,
\end{equation}
and if we only consider the recoil effect, it will become
\begin{equation}
 \Gamma_{cr}=\frac{\Gamma_{0}\omega_{0}}{m }\left[-\frac{5}{2}+\frac{3}{4}\left(\frac{Z^{2}+2}{Z^{2}}\cos Z+\frac{3Z^{2}-2}{Z^{3}}\sin Z\right)\right].
\end{equation}
Similarly, we find that the R\"{o}ntgen term-induced correction and the recoil-induced correction can be simply added to obtain the total correction to the spontaneous decay rate in this case [see Fig.~\ref{figure.1}(b)]. Moreover,  we can see that the recoil effect also decreases the decay rate, while the R\"{o}ntgen term increases the decay rate.

As for  the dipole transition moment $\textbf{d}=(d,0,0)$ is along the direction of motion ($x$-direction), $\langle f|\hat{\textbf{P}}\cdot [\hat{\textbf{B}}(\textbf{r},t)\times \textbf{d}]
+[\hat{\textbf{B}}(\textbf{r},t)\times \textbf{d}]\cdot \hat{\textbf{P}}|i\rangle=d\langle f|-i\partial_{y}\hat{B}^{-}_{z}+i\partial_{z}\hat{B}^{-}_{y}|i\rangle$.
 The resulted spontaneous decay rate is
\begin{eqnarray}\label{12}
  \Gamma&=& 2\pi d^{2}\sum_{\textbf{k}\lambda}\varepsilon^{2}_{k}F^{(3)}_{\lambda}(\omega_{\textbf{k}})
  \delta(\omega_0-\omega_{\textbf{k}}+k_{x}v)\sin^{2}(k_{z}z)\nonumber\\
  &=&\Gamma_{bx}-\frac{3\Gamma_{0}\omega_{0}}{2m}\bigg(1-\frac{\cos Z}{2}-\frac{\sin Z}{2Z}\bigg),
\end{eqnarray}
where $F^{(3)}_{\lambda}(\omega_{\textbf{k}})=|(e_{\textbf{k}\lambda})_{x}-h_{y}
\omega_{\textbf{k}}(\textbf{h}\times \textbf{e}_{\textbf{k}\lambda})_{z}/(2m)+h_{z}
\omega_{\textbf{k}}(\textbf{h}\times \textbf{e}_{\textbf{k}\lambda})_{y}/(2m)|^{2}$ and
 $\Gamma_{bx}=\Gamma_{by}$. Here we have used $\sum_{\lambda}F^{(3)}_{\lambda}(\omega_{\textbf{k}})=1-h^{2}_{x}+\left(h^{2}_{y}+h^{2}_{z}\right)
\omega_{\textbf{k}}/m$.
We plot the corresponding corrections for the case of the dipole along the $x$-direction in Fig.~\ref{figure.1}(c), which show the same features as those in Fig.~\ref{figure.1}(b) with the dipole along the $y$-direction. One notices
that, when the dipole transition moment is parallel to the plate,
the emission rate of the atom (fixed or not) are the same. Besides, we can find that
 the recoil-induced and the R\"{o}ntgen term-induced correction are the same with the ones
 when the dipole transition
moment is in $y$-direction [see Fig.~\ref{figure.1}(c)]. Thus,
the motion does not break the symmetry of the spontaneous emission rate for atom with the dipole in a plane parallel to the boundary.

In conclusion, we study the spontaneous emission rate of a moving atom near an infinitely perfectly-conducting plate. Different directions of the dipole moment of the atom have been considered. The corrections from the R\"{o}ntgen term and the recoil effect have been explored. It has been found that the total correction holds the additivity from individual contributions. The individual contributions induced solely by the R\"{o}ntgen term and the recoil effect are positive and negative, respectively. Moreover, it is interesting to note that the motion of atom does not break the equality of the spontaneous emission rate for the dipole in different directions but in the same plane parallel to the plate. Our work points towards interesting physics in quantum optics associated with the light-matter interaction for moving atoms in the future.

A.Z. would like to thank L. Yuan for revising this paper.




\end{document}